\documentclass[pra,showpacs,amsmath,amssymb,twocolumn,superscriptaddress]{revtex4}
\usepackage{bm,multirow}
\usepackage[usenames]{color}
\newcommand \beq{\begin{eqnarray}}
\newcommand \eeq{\end{eqnarray}}
\def\simge{\mathrel{%
       \rlap{\raise 0.511ex \hbox{$>$}}{\lower 0.511ex \hbox{$\sim$}}}}
\def\simle{\mathrel{\rlap{\raise 0.511ex \hbox{$<$}}{\lower 0.511ex \hbox{$\sim$}}}}
\usepackage{graphics}
\usepackage[dvips]{graphicx}
\usepackage{graphicx}

\newcommand{\be}{\begin{equation}}
\newcommand{\ee}{\end{equation}}
\newcommand{\bea}{\begin{eqnarray}}
\newcommand{\eea}{\end{eqnarray}}
\newcommand{\ba}{\begin{align}}
\newcommand{\ea}{\end{align}}

\newcommand{\rme}{{\rm e}}
\newcommand{\rmi}{{\rm i}}
\newcommand{\rmo}{{\rm o}}

\begin{document}
\title{Nucleus--nucleus interactions in the inner crust of neutron stars}
\author{D. Kobyakov}
\affiliation{Institute of Applied Physics of the Russian Academy of Sciences, 603950 Nizhny Novgorod, Russia
}
\affiliation{Institute of Nuclear Physics, Technical University Darmstadt,
D-64289 Darmstadt, Germany}

\author{C. J. Pethick}
\affiliation{The Niels Bohr International Academy, The Niels Bohr Institute, University of Copenhagen, Blegdamsvej 17, DK-2100 Copenhagen \O, Denmark}
\affiliation{NORDITA, KTH Royal Institute of Technology and Stockholm University, Roslagstullsbacken 23, SE-106 91 Stockholm, Sweden}

\date{\today}

\begin{abstract}
The interaction between nuclei in the inner crust of neutron stars consists of two contributions, the so-called ``direct'' interaction and an ``induced'' one due to density changes in the neutron fluid.   For large nuclear separations $r$ the contributions from nuclear forces to each of these terms are shown to be nonzero.  In the static limit they are equal in magnitude but have opposite signs and they cancel exactly.  We analyze earlier results on effective interactions in the light of this finding.   We consider the properties of long-wavelength collective modes and, in particular, calculate the degree of mixing between the lattice phonons and the phonons in the neutron superfluid.  Using microscopic theory, we calculate the net non-Coulombic contribution to the nucleus--nucleus interaction and show that, for large $r$, the leading term is due to exchange of two phonons and varies as $1/r^7$: it is an analog of the Casimir--Polder interaction between neutral atoms.

\pacs{ }

\end{abstract}

\maketitle
\section{Introduction}  In the inner crust of neutron stars, neutron-rich atomic nuclei coexist with a neutron fluid in addition to the background of electrons which ensures electrical neutrality.  Consequently, there are contributions to the effective interaction between nuclei due to the presence of the neutrons outside nuclei, in addition to the screened Coulomb interaction.    These interactions determine frequencies of collective modes of the crust \cite{Epstein, ChamelPageReddy,KP1} and it has been suggested that they could alter the equilibrium lattice structure \cite{KP2}.  In addition, they are an important ingredient in calculations of dynamical phenomena.

The approach in this article is to use thermodynamic reasoning, as was done in the case of dilute solutions of $^3$He in superfluid liquid $^4$He \cite{BBP1}.  An important difference between the present problem and that of liquid helium mixtures is that the translational kinetic energy of the nuclei is relatively unimportant and one may work in terms of an effective potential between nuclei that depends on their positions.  For helium mixtures, the fact that $^3$He atoms have a mass comparable to that of the $^4$He atoms means that it is more natural to work with  quantum-mechanical eigenstates in which the $^3$He atoms are not localized in space but are spread out over the whole of the system.

In many-body physics it  is common to express the effective interaction between particles as the sum of two terms, a `direct' interaction, which represents the energy change when the average particle density is held fixed and an `induced' interaction, which results from changes in the average density \cite{BabuBrown}.    In the thermodynamic approach to effective interactions we shall demonstrate that at large nuclear separations these contributions are nonzero, equal in magnitude but opposite in sign: thus they cancel exactly.

This paper is organized as follows.  In Sec.\ II we review the thermodynamic approach to effective interactions in two-component systems and describe its implications for the spatial dependence of the interactions.  We then analyse the results obtained from the calculations of Lattimer and Swesty \cite{LattimerSwesty} in the light of these findings.  Velocities of long-wavelength collective modes are calculated in Sec.\ III.  The formalism employed in Refs.\ \cite{ChamelPageReddy,KP1} is based on the traditional formulation of the two-fluid model for superfluidity \cite{LandLHydro, CJPChamelReddy}, in which one works with the velocity of the normal component (a contravariant vector) and the so-called superfluid velocity (a covariant vector), and we comment on how this is related to a treatment in terms of two covariant vector quantities, as is more natural in the context of a mixture of two superfluids, such as may be anticipated to occur in the
outer
core of neutron stars.  There are two longitudinal modes, which may be thought of as hybrids of lattice phonons and the Bogoliubov--Anderson mode of the neutron superfluid.  We give numerical results for the degree of hybridization, which is an important ingredient in calculations of the damping of modes.  We show that the hybridization  is rather sensitive to the neutron superfluid density, a quantity whose magnitude is rather uncertain.  Section IV contains a microscopic calculation of the interaction between nuclei and we show that the long-range part of the interaction is attractive and varies as $1/r^7$ \cite{Schecter}.     Section V contains concluding remarks, and in the Appendix we give an explicit example of the cancellation of direct and induced interactions in a model of matter in the inner crust in which Coulomb and surface effects are neglected.

\section{Basic considerations}

Locally, the densities of nucleons are inhomogeneous because of the presence of the nuclei and it is convenient to work in terms of the average densities of neutrons, $n_n$, and of protons, $n_p$.   We shall assume that matter is electrically neutral and therefore the electron density $n_e$ is equal to $n_p$.  In the inner crust, protons are all in nuclei and therefore the density of nuclei is given by $n_N=n_e/Z$, where $Z$ is the proton number of the nucleus.  We shall examine how the energy density $E(n_n,n_p)$ depends on the densities of neutrons and protons, and in making variations, we shall assume that $Z$ remains constant.  To change the atomic number of a nucleus at low temperature is a slow process, since it requires a major reorganization of the protons:  either a nucleus must disappear by its individual protons being absorbed by other nuclei, or a nucleus can be created by, say clustering of one or more protons removed from each of a number of initial nuclei.  Such processes are strongly inhibited by the high energy barriers required to eject many protons from nuclei.  In addition, calculations indicate that the equilibrium value of $Z$ in the inner crust is rather insensitive to density
\cite{RavenhallBP,NegeleVautherin,DouchinHaensel,GrillMargueronSandulescu,PearsonChamel,BaldoBurgio}. It is 
therefore reasonable to assume that nuclei do not appear or disappear except on very long timescales.
We shall also neglect weak interaction processes, since the timescale for establishment of thermal equilibrium is short compared with weak-interaction timescales.

For a binary mixture consisting of two components $1$ and $2$, the condition for the system to be stable to small density inhomogeneities is positive definiteness of the second order perturbation of energy,
\beq
\delta^2E=&\frac12 \sum_{i,j}E_{ij}\delta n_i\delta n_j>0,
\label{E2}
\eeq
or
\beq
E_{11}>0\,\,\,,E_{22}>0,
\label{instab1}
\eeq
and
\beq
E_{11}E_{22}-E_{12}^2>0,
\label{instab2}
\eeq
where $E_{ij}=\partial^2E/\partial n_i\partial n_j$, $E(n_i,n_j)$ being the energy density of the system in terms of the number densities, $n_i$ and $n_j$, of the two components.  Sufficient conditions for stability are that inequality (\ref{instab2}) and
one of the inequalities (\ref{instab1}) are satisfied, since if that is the case, the second of the inequalities (\ref{instab1}) is automatically satisfied.

For matter in the inner crust and for liquid helium mixtures, the diagonal terms $E_{11}$ and $E_{22}$ are positive, so inequality (\ref{instab2}) may be written as
\beq
E_{22}-\frac{E_{12}^2}{E_{11}}>0.
\label{instab3}
\eeq
The quantity $E_{22}$ in inequality (\ref{instab3}) is the direct interaction and it represents energy changes when $n_2$ is changed, keeping $n_1$ fixed.  The induced interaction
\be
V_{22}^{\rm ind}=-\frac{E_{12}^2}{E_{11}}
\label{indint}
\ee
represents the energy reduction due to changes in  $n_1$.   This has the expected form for such an interaction:  the square of the coupling between the two components, $E_{12}$, multiplied by the density--density response function for component $1$, which in the long-wavelength limit is given by $-\partial n_1/\partial \mu_1=-1/E_{11}$.  The induced interaction may be rewritten in a form familiar from the theory of dilute solutions of $^3$He in liquid $^4$He as \cite{BBP1}
\be
V_{22}^{\rm ind}=-\nu^2 E_{11}=-\nu^2\frac{\partial \mu_1}{\partial n_1},
\label{indint2}
\ee
where
\be
\nu=-E_{12}/E_{11}=\left.\partial n_1/\partial n_2\right|_{\mu_1}
\label{nu}
\ee
 is the number of component $1$ added when a single ``particle'' of species $2$ is added at constant $\mu_1$.

The left hand side of (\ref{instab3}) may be written more compactly
since
\beq
E_{22}-\frac{E_{12}^2}{E_{11}}=\left.\frac{\partial \mu_2}{\partial n_2}\right|_{\mu_1},
\label{identity}
\eeq
where
the chemical potentials of the components are given by
\beq
\mu_i=\frac{\partial E}{\partial n_i}.
\eeq
The identity (\ref{identity}) follows directly from the fact that
\beq
\left.\frac{\partial \mu_j}{\partial n_j}\right|_{\mu_i}=\frac{\partial \mu_j}{\partial n_j} +\frac{\partial \mu_j}{\partial n_i}\left.\frac{\partial n_i}{\partial n_j}\right|_{\mu_i}
\eeq
and
\beq
\left.\frac{\partial n_i}{\partial n_j}\right|_{\mu_i}=-\frac{\partial \mu_i}{\partial n_j}\Big/\frac{\partial \mu_i}{\partial n_i}=-\frac{E_{ij}}{E_{ii}}.
\eeq

An advantage of working in terms of $\mu_i$ rather than $n_i$ is that the second order change in the energy density is given by
\beq
\delta^2E=&\frac12 \frac{\partial n_1}{\partial \mu_1} (\delta \mu_1)^2 +\frac12 \left.\frac{\partial \mu_2}{\partial n_2}\right|_{\mu_1}(\delta n_2)^2,
\label{d2E}
\eeq
which has no cross terms.
\subsection{Spatial dependence of interaction}

In studying dilute solutions of $^3$He in liquid $^4$He it is customary to work in terms of states in which $^3$He atoms are in states that are spread throughout the volume of the system.  For applications to nuclei in neutron stars, the mass of the nucleus is typically 100 or more times that of a neutron, and the kinetic energy associated with the translational motion of the nucleus is unimportant.  Consequently it is meaningful to consider the interaction between nuclei with definite positions and we now examine the induced interaction from this viewpoint.

As a simple example, let us consider the interaction between two voids (component 2) of volume $v_2$ in a quantum liquid, component 1.  For the moment we imagine the particles in the liquid to be uncharged and we shall consider the effects of the Coulomb interaction later.  We shall also neglect the surface energy associated with the void.  When one void is created in the liquid, keeping the total number of particles  and the volume of the system fixed, the fractional increase in the density of the liquid over the whole of space except for a region in the immediate vicinity of the void is approximately $v_2/{\cal V}$, where ${\cal V}$ is the volume of the system.  The energy to create a second void even very far from the first one will be greater than that to create the first one since the density of the liquid even far from the first void is greater than when there is no void.  The extra energy required to add the second void compared with the first one is the direct interaction between the two voids.

For a simple quantitative model, we calculate the energy of the system by imagining that the effect of adding a void is simply to reduce the volume available to the liquid by an amount $v_2$.  Thus the volume available to the liquid is ${\cal V}-N_2v_2$ and the total energy $\cal E$ of the system with $N_2$ voids is given by
\beq
{\cal E}(N_1, N_2) =({\cal V}-N_2v_2)E^0\left(\frac{N_1}{{\cal V}-N_2v_2}\right),
\eeq
where $E^0(N_1/{\cal V})$ is the energy density of the liquid with no voids. The energy of interaction between two voids in a fluid of $1$-particles is
\beq
{\cal E}(N_1, 2)-{\cal E}(N_1, 0) -2\left[{\cal E}(N_1, 1)-{\cal E}(N_1, 0) \right]\nonumber\\ ={\cal E}(N_1, 2)-2{\cal E}(N_1, 1)+{\cal E}(N_1, 0)\approx \frac{(n_1v_2)^2}{{\cal V}} \frac{\partial \mu_1}{\partial n_1},
\eeq
which, when expressed in terms of energy densities and particle densities, is
\beq
E_{22}= (n_1v_2)^2 \frac{\partial \mu_1}{\partial n_1}.
\eeq
When the positions of the voids are regarded as fixed, there are no kinetic contributions to $E_{22}$, which is thus the direct contribution to the effective interaction.

We now turn to the induced interaction,  Eq.\ (\ref{indint2}).   The quantity $\nu$, Eq.\ (\ref{nu}) is $-n_1v_2$, because addition of a void at constant $\mu_1$ results in the loss of $n_1v_2$ 1-particles in the volume of the void \cite{1+alpha}.  Thus one finds
\be
V_{22}^{\rm ind}=-(n_1v_2)^2 E_{11}.
\ee
Consequently
the induced interaction precisely cancels the direct interaction in this model,
\be
E_{22}+V_{22}^{\rm ind}=0.
\label{instab5}
\ee

 \begingroup
\squeezetable
\begin{table*}
\caption{Microscopic parameters of crustal matter. The quantity $n^{\rm o}$ is the neutron density outside nuclei, $x^{\rm in}$ is the proton fraction of matter in the interior of nuclei, $u$ is the volume fraction of a nucleus in a unit cell, $A$ ($Z$) is the number of nucleons (protons) in a nucleus.}
\label{tab1}
\begin{center}
\begin{tabular}{|c|c|c|c|c|c|c|c|c|c|c|c|c|}
  \hline
  $n [\texttt{fm}^{-3} ]$&$n^{\rm o}  [\texttt{fm}^{-3} ]$&$\mu_e  [\texttt{MeV}]$&$Y_e=n_p/n$&$x^{\rm in}$ &$u$&$A$&$Z$&$E_{nn}\,[\texttt{GeV}\,\texttt{fm}^3] $&$E_{np}\,[\texttt{GeV}\,\texttt{fm}^3] $&$E_{pp}\,[\texttt{GeV}\,\texttt{fm}^3] $&$\frac{\partial\mu_e}{\partial n_e} \,[\texttt{GeV}\,\texttt{fm}^3]$&$\Gamma $\\ \hline

2.512$\times10^{-4}$ & 1.138$\times10^{-6}$ & 27.29 & 0.3557 & 0.3573 & 1.380$\times10^{-3}$ & 106.0 & 37.87 & 16.11 & -33.37 & 148.2 & 101.8 & 1.338\\
2.818$\times10^{-4}$ & 1.877$\times10^{-5}$ & 27.72 & 0.3320 & 0.3557 & 1.451$\times10^{-3}$ & 107.0 & 38.05 & 4.820 & -12.69 & 109.3 & 98.70 & 2.128\\
3.162$\times10^{-4}$ & 4.239$\times10^{-5}$ & 28.06 & 0.3070 & 0.3544 & 1.510$\times10^{-3}$ & 107.8 & 38.20 & 3.467 & -9.829 & 102.9 & 96.32 & 2.403\\
3.548$\times10^{-4}$ & 7.051$\times10^{-5}$ & 28.38 & 0.2832 & 0.3532 & 1.567$\times10^{-3}$ & 108.5 & 38.33 & 2.822 & -8.284 & 98.98 & 94.14 & 2.503\\
3.981$\times10^{-4}$ & 1.032$\times10^{-4}$ & 28.70 & 0.2610 & 0.3521 & 1.626$\times10^{-3}$ & 109.3 & 38.47 & 2.425 & -7.228 & 95.87 & 92.06 & 2.508\\
4.467$\times10^{-4}$ & 1.407$\times10^{-4}$ & 29.03 & 0.2405 & 0.3509 & 1.686$\times10^{-3}$ & 110.0 & 38.61 & 2.146 & -6.430 & 93.16 & 90.03 & 2.462\\
5.012$\times10^{-4}$ & 1.836$\times10^{-4}$ & 29.36 & 0.2218 & 0.3496 & 1.750$\times10^{-3}$ & 110.8 & 38.75 & 1.934 & -5.791 & 90.66 & 88.00 & 2.390\\
5.623$\times10^{-4}$ & 2.323$\times10^{-4}$ & 29.70 & 0.2047 & 0.3484 & 1.818$\times10^{-3}$ & 111.6 & 38.89 & 1.763 & -5.262 & 88.29 & 85.98 & 2.306\\
6.310$\times10^{-4}$ & 2.877$\times10^{-4}$ & 30.06 & 0.1891 & 0.3470 & 1.891$\times10^{-3}$ & 112.5 & 39.04 & 1.621 & -4.814 & 85.98 & 83.95 & 2.220\\
7.079$\times10^{-4}$ & 3.504$\times10^{-4}$ & 30.43 & 0.1749 & 0.3457 & 1.969$\times10^{-3}$ & 113.4 & 39.20 & 1.499 & -4.427 & 83.71 & 81.90 & 2.135\\
7.943$\times10^{-4}$ & 4.213$\times10^{-4}$ & 30.83 & 0.1620 & 0.3442 & 2.054$\times10^{-3}$ & 114.4 & 39.37 & 1.392 & -4.089 & 81.45 & 79.82 & 2.055\\
8.913$\times10^{-4}$ & 5.014$\times10^{-4}$ & 31.24 & 0.1503 & 0.3427 & 2.147$\times10^{-3}$ & 115.4 & 39.55 & 1.296 & -3.790 & 79.20 & 77.72 & 1.980\\
1.000$\times10^{-3}$ & 5.918$\times10^{-4}$ & 31.68 & 0.1397 & 0.3411 & 2.248$\times10^{-3}$ & 116.5 & 39.73 & 1.209 & -3.523 & 76.94 & 75.58 & 1.912\\
1.122$\times10^{-3}$ & 6.938$\times10^{-4}$ & 32.14 & 0.1300 & 0.3394 & 2.359$\times10^{-3}$ & 117.7 & 39.93 & 1.129 & -3.284 & 74.67 & 73.42 & 1.849\\
1.259$\times10^{-3}$ & 8.087$\times10^{-4}$ & 32.63 & 0.1213 & 0.3376 & 2.480$\times10^{-3}$ & 118.9 & 40.14 & 1.056 & -3.067 & 72.39 & 71.23 & 1.793\\
1.413$\times10^{-3}$ & 9.382$\times10^{-4}$ & 33.15 & 0.1133 & 0.3357 & 2.614$\times10^{-3}$ & 120.2 & 40.36 & 0.9888 & -2.870 & 70.10 & 69.02 & 1.741\\
1.585$\times10^{-3}$ & 1.084$\times10^{-3}$ & 33.70 & 0.1061 & 0.3336 & 2.762$\times10^{-3}$ & 121.7 & 40.60 & 0.9259 & -2.691 & 67.80 & 66.78 & 1.695\\
1.778$\times10^{-3}$ & 1.248$\times10^{-3}$ & 34.28 & 0.09957 & 0.3315 & 2.926$\times10^{-3}$ & 123.2 & 40.85 & 0.8672 & -2.525 & 65.49 & 64.53 & 1.653\\
1.995$\times10^{-3}$ & 1.432$\times10^{-3}$ & 34.90 & 0.09365 & 0.3292 & 3.107$\times10^{-3}$ & 124.9 & 41.11 & 0.8121 & -2.373 & 63.17 & 62.25 & 1.615\\
2.239$\times10^{-3}$ & 1.639$\times10^{-3}$ & 35.56 & 0.08827 & 0.3268 & 3.309$\times10^{-3}$ & 126.7 & 41.40 & 0.7604 & -2.232 & 60.84 & 59.97 & 1.580\\
2.512$\times10^{-3}$ & 1.872$\times10^{-3}$ & 36.26 & 0.08340 & 0.3243 & 3.534$\times10^{-3}$ & 128.6 & 41.70 & 0.7117 & -2.102 & 58.52 & 57.68 & 1.549\\
2.818$\times10^{-3}$ & 2.134$\times10^{-3}$ & 37.00 & 0.07899 & 0.3216 & 3.785$\times10^{-3}$ & 130.7 & 42.02 & 0.6658 & -1.980 & 56.21 & 55.39 & 1.521\\
3.162$\times10^{-3}$ & 2.428$\times10^{-3}$ & 37.79 & 0.07499 & 0.3187 & 4.066$\times10^{-3}$ & 132.9 & 42.35 & 0.6225 & -1.866 & 53.91 & 53.11 & 1.495\\
3.548$\times10^{-3}$ & 2.758$\times10^{-3}$ & 38.62 & 0.07136 & 0.3156 & 4.382$\times10^{-3}$ & 135.3 & 42.71 & 0.5815 & -1.759 & 51.62 & 50.84 & 1.472\\
3.981$\times10^{-3}$ & 3.128$\times10^{-3}$ & 39.51 & 0.06807 & 0.3124 & 4.737$\times10^{-3}$ & 137.9 & 43.09 & 0.5428 & -1.659 & 49.36 & 48.59 & 1.451\\
4.467$\times10^{-3}$ & 3.544$\times10^{-3}$ & 40.45 & 0.06509 & 0.3090 & 5.137$\times10^{-3}$ & 140.8 & 43.50 & 0.5061 & -1.564 & 47.12 & 46.36 & 1.430\\
5.012$\times10^{-3}$ & 4.010$\times10^{-3}$ & 41.44 & 0.06239 & 0.3054 & 5.589$\times10^{-3}$ & 143.8 & 43.92 & 0.4714 & -1.475 & 44.92 & 44.17 & 1.414\\
5.623$\times10^{-3}$ & 4.533$\times10^{-3}$ & 42.49 & 0.05994 & 0.3016 & 6.101$\times10^{-3}$ & 147.1 & 44.37 & 0.4385 & -1.391 & 42.76 & 42.01 & 1.398\\
6.310$\times10^{-3}$ & 5.120$\times10^{-3}$ & 43.60 & 0.05772 & 0.2975 & 6.681$\times10^{-3}$ & 150.7 & 44.85 & 0.4073 & -1.311 & 40.65 & 39.90 & 1.383\\
7.079$\times10^{-3}$ & 5.777$\times10^{-3}$ & 44.77 & 0.05571 & 0.2933 & 7.340$\times10^{-3}$ & 154.6 & 45.34 & 0.3778 & -1.235 & 38.59 & 37.84 & 1.369\\
7.943$\times10^{-3}$ & 6.514$\times10^{-3}$ & 46.01 & 0.05388 & 0.2888 & 8.092$\times10^{-3}$ & 158.8 & 45.87 & 0.3499 & -1.163 & 36.58 & 35.83 & 1.357\\
8.913$\times10^{-3}$ & 7.340$\times10^{-3}$ & 47.31 & 0.05221 & 0.2841 & 8.949$\times10^{-3}$ & 163.4 & 46.41 & 0.3235 & -1.094 & 34.64 & 33.89 & 1.346\\
1.000$\times10^{-2}$ & 8.266$\times10^{-3}$ & 48.68 & 0.05069 & 0.2791 & 9.931$\times10^{-3}$ & 168.4 & 46.98 & 0.2986 & -1.028 & 32.76 & 32.01 & 1.335\\
1.122$\times10^{-2}$ & 9.303$\times10^{-3}$ & 50.11 & 0.04930 & 0.2738 & 1.106$\times10^{-2}$ & 173.7 & 47.57 & 0.2751 & -0.9657 & 30.95 & 30.20 & 1.325\\
1.259$\times10^{-2}$ & 1.047$\times10^{-2}$ & 51.62 & 0.04802 & 0.2683 & 1.235$\times10^{-2}$ & 179.5 & 48.18 & 0.2531 & -0.9060 & 29.21 & 28.46 & 1.316\\
1.413$\times10^{-2}$ & 1.177$\times10^{-2}$ & 53.20 & 0.04684 & 0.2626 & 1.384$\times10^{-2}$ & 185.8 & 48.79 & 0.2324 & -0.8491 & 27.54 & 26.80 & 1.307\\
1.585$\times10^{-2}$ & 1.323$\times10^{-2}$ & 54.85 & 0.04575 & 0.2565 & 1.556$\times10^{-2}$ & 192.7 & 49.42 & 0.2131 & -0.7950 & 25.95 & 25.21 & 1.299\\
1.778$\times10^{-2}$ & 1.486$\times10^{-2}$ & 56.57 & 0.04473 & 0.2502 & 1.755$\times10^{-2}$ & 200.0 & 50.04 & 0.1951 & -0.7437 & 24.43 & 23.70 & 1.291\\
1.995$\times10^{-2}$ & 1.670$\times10^{-2}$ & 58.36 & 0.04378 & 0.2435 & 1.987$\times10^{-2}$ & 207.9 & 50.64 & 0.1786 & -0.6953 & 22.99 & 22.27 & 1.284\\
2.239$\times10^{-2}$ & 1.875$\times10^{-2}$ & 60.23 & 0.04289 & 0.2366 & 2.257$\times10^{-2}$ & 216.4 & 51.20 & 0.1634 & -0.6500 & 21.62 & 20.91 & 1.276\\
2.512$\times10^{-2}$ & 2.106$\times10^{-2}$ & 62.17 & 0.04204 & 0.2293 & 2.575$\times10^{-2}$ & 225.5 & 51.70 & 0.1496 & -0.6079 & 20.33 & 19.62 & 1.268\\
2.818$\times10^{-2}$ & 2.364$\times10^{-2}$ & 64.19 & 0.04124 & 0.2216 & 2.950$\times10^{-2}$ & 235.0 & 52.08 & 0.1373 & -0.5694 & 19.11 & 18.41 & 1.260\\
3.162$\times10^{-2}$ & 2.653$\times10^{-2}$ & 66.29 & 0.04049 & 0.2136 & 3.397$\times10^{-2}$ & 244.9 & 52.30 & 0.1266 & -0.5349 & 17.96 & 17.26 & 1.251\\
3.548$\times10^{-2}$ & 2.977$\times10^{-2}$ & 68.49 & 0.03980 & 0.2051 & 3.936$\times10^{-2}$ & 254.9 & 52.26 & 0.1174 & -0.5050 & 16.88 & 16.17 & 1.243\\
3.981$\times10^{-2}$ & 3.339$\times10^{-2}$ & 70.79 & 0.03917 & 0.1960 & 4.596$\times10^{-2}$ & 264.4 & 51.82 & 0.1099 & -0.4802 & 15.87 & 15.13 & 1.234\\
4.467$\times10^{-2}$ & 3.743$\times10^{-2}$ & 73.22 & 0.03863 & 0.1862 & 5.421$\times10^{-2}$ & 272.6 & 50.76 & 0.1041 & -0.4611 & 14.93 & 14.14 & 1.225\\
5.012$\times10^{-2}$ & 4.193$\times10^{-2}$ & 75.81 & 0.03820 & 0.1756 & 6.480$\times10^{-2}$ & 277.4 & 48.71 & 0.1001 & -0.4484 & 14.04 & 13.20 & 1.218\\
5.623$\times10^{-2}$ & 4.693$\times10^{-2}$ & 78.57 & 0.03791 & 0.1639 & 7.884$\times10^{-2}$ & 275.3 & 45.12 & 0.09805 & -0.4426 & 13.22 & 12.28 & 1.214\\
6.310$\times10^{-2}$ & 5.244$\times10^{-2}$ & 81.55 & 0.03777 & 0.1507 & 9.828$\times10^{-2}$ & 259.8 & 39.16 & 0.09787 & -0.4438 & 12.46 & 11.40 & 1.215\\
7.079$\times10^{-2}$ & 5.849$\times10^{-2}$ & 84.76 & 0.03780 & 0.1357 & 1.267$\times10^{-1}$ & 220.2 & 29.88 & 0.09963 & -0.4516 & 11.77 & 10.56 & 1.222\\
7.943$\times10^{-2}$ & 6.510$\times10^{-2}$ & 88.19 & 0.03795 & 0.1183 & 1.712$\times10^{-1}$ & 145.6 & 17.23 & 0.1036 & -0.4635 & 11.14 & 9.751 & 1.227\\
8.913$\times10^{-2}$ & 7.241$\times10^{-2}$ & 91.78 & 0.03813 & 0.09828 & 2.466$\times10^{-1}$ & 45.72 & 4.493 & 0.1120 & -0.4747 & 10.59 & 9.003 & 1.187\\

 \hline
\end{tabular}
\end{center}
\end{table*}
\endgroup

\squeezetable

In the simple model of voids in a quantum liquid, the cancellation of the direct and induced interactions is easily understood by noting that the effective interaction represents the extra energy required to add, at fixed $\mu_1$, a second void compared  with that to add the first one (cf. Eq.\ (\ref{d2E})).  Since the energy of a void depends only on $\mu_1$ in this model, irrespective of the number of voids, the total energy of the system is linear in the number of voids and consequently the total interaction of any pair of voids, $\partial \mu_2/\partial n_2|_{\mu_1}$, vanishes.
\subsection{Nuclei in a neutron fluid}
We turn now to the case of nuclei in a neutron fluid, and we begin by neglecting the Coulomb interaction and the charge-neutralizing background of electrons.  The most widely used models of nuclei in the inner crust of neutron stars are based on the liquid drop picture, with the energy given as a sum of a bulk energy of matter inside nuclei, a bulk energy of the neutrons outside nuclei, and a surface contribution.  For such models it is immediately apparent that the total effective interaction between nuclei vanishes, provided the nuclei do not overlap.  The result is even more general, since it does not depend on how the interior of the nucleus is described:  the only requirement on the model is that the energy of the neutron fluid outside may be treated as a uniform fluid with chemical potential $\mu_1$.  In Appendix A we derive explicit expressions for the thermodynamic derivatives when surface effects are neglected and show that condition (\ref{instab5}) is satisfied.

We now investigate the results of calculations of the quantities $E_{ij}$ for matter in the inner crust in the light of these findings.  The approach of Lattimer and Swesty \cite{LattimerSwesty} is based on a liquid drop model and therefore the direct and induced interactions should satisfy the identity (\ref{instab5}) if Coulomb effects are neglected.  In the calculations of Ref.\ \cite{LattimerSwesty}, Coulomb effects enter in a number of places. First, they ensure that the average electron and proton densities are equal, and thus the electron density is  changed when the proton density is altered.  This contributes to $E_{pp}$ an amount
\be
(E_{pp})_{e}=\frac{\partial^2 E_e}{\partial n_e^2}=\frac{\partial \mu_e}{\partial n_e}\approx \frac13 \frac{\mu_e}{n_e}.
\label{electronbulkmod}
\ee
Here $E_e$ is the energy density of the non-interacting electron gas and $n_e$ the electron number density.  The final result in Eq.\ (\ref{electronbulkmod}) is for ultrarelativistic electrons, for which $E_e\propto n_e^{4/3}$.
Second, the Coulomb energy of the lattice of nuclei changes.  In the approximation that the unit cell may be taken to be spherical, the Coulomb energy of the lattice per unit volume is
\be
E^{\rm latt}=-\frac9{10}n_N\frac{Z^2e^2}{r_c}= -\frac9{10}n_e\frac{Z^{2/3}e^2}{r_e},
\ee
where $n_N$ is the number density of nuclei and $r_c$ is defined by the equation $\frac{4}{3}\pi r_c^3n_N=1$ and $r_e$ by $\frac{4}{3}\pi r_e^3n_e=1$.  The Coulomb energy also contributes to the energy of a single nucleus immersed in the neutron fluid.  The electrostatic energy of the protons in a nucleus interacting with themselves gives a contribution to the energy which is linear in the number of nuclei and it therefore does not contribute to $E_{pp}$.  There is a correction to the Coulomb lattice energy due to the fact that the proton distribution in a nucleus is extended rather than point-like  and this does depend on the density of nuclei.   However this is a numerically small effect except at densities close to the crust--core boundary and we shall neglect it.

Since the chemical potential of degenerate, ultrarelativistic  electrons is given by $\mu_e=\hbar ck_e$, where $k_e=(3\pi^2n_e)^{1/3}$ is the electron Fermi wave number, it follows the  ratio of the lattice and electronic contributions to $E_{pp}$ is given by
\be
\frac{E^{\rm latt}}{E_{pp}^{e}}= -\frac{2}{5}  \left(\frac{12}{\pi} \right)^{1/3} Z^{2/3}\frac{e^2}{\hbar c}\approx      -0.05337
\left(\frac{Z}{40}\right)^{2/3}.
\ee
We now examine the results of Ref.\ \cite{KP1}.  These are shown in Table 1, where we have corrected a numerical error in the original paper.
To estimate the contribution of nuclear forces to the thermodynamic derivative, we must remove the effects of the Coulomb interaction.  These are greatest for $E_{pp}$, because the Coulomb interaction forces the electron density to be equal to the proton density, thereby giving the dominant contribution $\partial \mu_e/\partial n_e$ to $E_{pp}$.  In addition, the energy of the Coulomb lattice of nuclei contributes  $ -0.05337
\left({Z}/{40}\right)^{2/3} \partial \mu_e/\partial n_e$ to $E_{pp}$.  It is more difficult to make simple estimates of the Coulomb contributions to $E_{np}$, but we expect these to be relatively small since the Coulomb energy per proton in a nucleus is small compared with the proton chemical potential, $\sim -50$ MeV \cite{mue included in Epp}.

As an example, we consider the case of a density of $10^{-2}$ fm$^{-3}$.  We estimate the contribution to $E_{pp}$ from nuclear forces by removing the electronic and lattice contributions and find
\begin{align}
E_{pp}^{\rm nuc}&=E_{pp}- \frac{\partial \mu_e}{\partial n_e}\left(1-0.05337
\left(\frac{Z}{40}\right)^{2/3}\right)\nonumber \\
&=  32760-32010 [ 1-0.05337(46.48/40)^{2/3}] ~ \rm {MeV~fm}^3 \nonumber\\
&=    2652  ~{\rm MeV~fm}^3.
\end{align}
Extracting the nuclear contributions to $E_{np}$ and $E_{nn}$ from results for the calculations of Ref.~\cite{LattimerSwesty} is a more difficult task, but we expect the effect of turning off the Coulomb interaction to be much less dramatic than for $E_{pp}$.  We therefore compare $E_{pp}^{\rm nuc}$ with the magnitude of the induced interaction calculated without removal of Coulomb effects, $E_{np}^2/E_{nn}$, and define the quantity
\be
\Gamma=\frac{E_{np}^2}{E_{pp}^{\rm nuc}E_{nn}}.
\ee
If the direct and induced interactions exactly cancelled, this ratio would be unity, and for the case considered here, it is 1.335, indicating a large degree of cancellation between contributions. We attribute the fact that this ratio is not unity to the fact that Coulomb contributions have not been removed from  $E_{np}$ and $E_{nn}$.
The parameter $\Gamma$ is calculated for other values of the baryon number density in Table I, and the results are of the same order of magnitude across the inner crust.

\section{Collective modes}
In Ref.\ \cite{KP1} we calculated velocities of collective modes in terms of the interactions between nucleons and the superfluid mass density $n_n^s$ of the neutrons.  The formalism  presented there is based on the standard treatment of the two-fluid model \cite{LandLHydro, CJPChamelReddy}, and the basic variables are the densities of protons and neutrons, the displacement of an ion (and the corresponding velocity, ${\bf v}_i$) and the so-called superfluid velocity, ${\bf v}^{s}=\hbar \boldsymbol{\nabla}\phi_n/m$, where $2\phi_n$ is the phase of the superfluid order parameter.    As has been stressed by various authors, the  two velocity variables have different characters, since   ${\bf v}_i$ is a contravariant vector while  ${\bf v}^{s}$ is a covariant vector \cite{Carter}.     A great advantage of this treatment is that it makes obvious from the outset that just one additional density variable, the superfluid density, is required to characterize the superfluid state.  In the core of neutron stars, where both neutrons and protons may be superfluid, it is natural to use a more symmetrical approach, in which one works with the gradients of the phases of the superfluid order parameters of both neutrons and protons.   The equivalence of the two approaches will be demonstrated explicitly elsewhere \cite{Kobyakov_etal}.

We now discuss the physical nature of the modes following Ref.\ \cite{KP1}.
When the nuclei are stationary,  the velocity of sound in the superfluid neutrons is given by
\beq
v_n^2=\frac{n_n^sE_{nn}}{m},
\label{vn}
\eeq
where $n_n^s$ is the neutron superfluid number density. When the superfluid is stationary, the velocity of longitudinal lattice phonons is given by
\beq
v_p^2=\frac{\mathcal{E}^{nn}+\frac43 S}{\rho^n}.
\label{vp}
\eeq
Here we have included the effects of the shear rigidity of the lattice.  We assume that the lattice is polycrystalline with crystallites having a size small compared with the wavelength of the mode and, consequently, the shear elastic properties are isotropic, and we denote the shear elastic constant by $S$.  The normal mass density $\rho^n=m(n_p+n_n^n)$ is the total normal mass density \cite{Releffects}, where $n_n^n=n_n-n_n^s$.  More generally, these two modes are coupled and their velocities $v=\omega/q$ satisfy the equation
\beq
(v^2-v_n^2)(v^2-v_p^2)-v_{np}^4=0,
\label{modevel}
\eeq
where the strength of the hybridization of the two modes is determined by
\beq
 v_{np}^2=\left(\frac{n_n^s}{m\rho^n}\right)^{1/2}(E_{nn}n_n^n+E_{np}n_p).
\label{vnp}
\eeq
There is significant cancellation in the two terms in the expression for $v_{np}^2$ since $E_{np}$ is negative and typically lies between $-2E_{nn}$ and $-5E_{nn}$: as a consequence, hybridization is weak.
Explicitly, the mode velocities are given by
\beq
v_{\pm}^2=\frac{v_n^2+v_p^2}2 \pm\sqrt{\left(\frac{v_n^2-v_p^2}{2}\right)^2+  v_{np}^4}.
\label{vpm}
\eeq
\begingroup
\squeezetable
\begin{table*}
\caption{Velocities of coupled and uncoupled modes $v_\pm$, $v_p^2$, $v_n^2$ and $v_{pn}^2$.}
\label{tab2}
\begin{center}
\begin{tabular}{|c|c|c|c|c|c|c|}
  \hline
  $n [\texttt{fm}^{-3} ]$&$v_+/c$ &$v_-/c$ & $v_p^2/c^2$ & $v_n^2/c^2$ & $v_{np}^2/c^2$ & $v_{np}^2/(v_p^2-v_n^2)$\\ \hline

2.512$\times10^{-4}$ & 5.283 & 0.4386 & 27.91 & 0.1952 & 0.2813 & 0.01015\\
2.818$\times10^{-4}$ & 5.359 & 0.9607 & 28.68 & 0.9629 & 1.053 & 0.03797\\
3.162$\times10^{-4}$ & 5.433 & 1.221 & 29.45 & 1.564 & 1.430 & 0.05127\\
3.548$\times10^{-4}$ & 5.501 & 1.421 & 30.16 & 2.118 & 1.661 & 0.05922\\
3.981$\times10^{-4}$ & 5.564 & 1.596 & 30.84 & 2.663 & 1.811 & 0.06428\\
4.467$\times10^{-4}$ & 5.621 & 1.756 & 31.47 & 3.213 & 1.911 & 0.06763\\
5.012$\times10^{-4}$ & 5.675 & 1.908 & 32.07 & 3.778 & 1.978 & 0.06993\\
5.623$\times10^{-4}$ & 5.725 & 2.053 & 32.64 & 4.361 & 2.025 & 0.07162\\
6.310$\times10^{-4}$ & 5.773 & 2.194 & 33.18 & 4.964 & 2.059 & 0.07296\\
7.079$\times10^{-4}$ & 5.819 & 2.332 & 33.71 & 5.590 & 2.084 & 0.07413\\
7.943$\times10^{-4}$ & 5.864 & 2.466 & 34.23 & 6.240 & 2.106 & 0.07526\\
8.913$\times10^{-4}$ & 5.907 & 2.599 & 34.73 & 6.914 & 2.126 & 0.07643\\
1.000$\times10^{-3}$ & 5.950 & 2.729 & 35.24 & 7.614 & 2.147 & 0.07773\\
1.122$\times10^{-3}$ & 5.992 & 2.858 & 35.74 & 8.340 & 2.170 & 0.07919\\
1.259$\times10^{-3}$ & 6.035 & 2.986 & 36.24 & 9.093 & 2.195 & 0.08086\\
1.413$\times10^{-3}$ & 6.077 & 3.113 & 36.75 & 9.873 & 2.225 & 0.08277\\
1.585$\times10^{-3}$ & 6.120 & 3.239 & 37.27 & 10.68 & 2.258 & 0.08494\\
1.778$\times10^{-3}$ & 6.164 & 3.364 & 37.79 & 11.52 & 2.297 & 0.08741\\
1.995$\times10^{-3}$ & 6.208 & 3.488 & 38.33 & 12.38 & 2.341 & 0.09020\\
2.239$\times10^{-3}$ & 6.253 & 3.612 & 38.88 & 13.27 & 2.390 & 0.09332\\
2.512$\times10^{-3}$ & 6.300 & 3.735 & 39.45 & 14.18 & 2.446 & 0.09679\\
2.818$\times10^{-3}$ & 6.347 & 3.857 & 40.03 & 15.12 & 2.507 & 0.1006\\
3.162$\times10^{-3}$ & 6.395 & 3.977 & 40.63 & 16.09 & 2.575 & 0.1049\\
3.548$\times10^{-3}$ & 6.445 & 4.097 & 41.25 & 17.07 & 2.649 & 0.1095\\
3.981$\times10^{-3}$ & 6.496 & 4.215 & 41.89 & 18.07 & 2.729 & 0.1146\\
4.467$\times10^{-3}$ & 6.548 & 4.331 & 42.54 & 19.09 & 2.816 & 0.1201\\
5.012$\times10^{-3}$ & 6.601 & 4.445 & 43.21 & 20.12 & 2.910 & 0.1260\\
5.623$\times10^{-3}$ & 6.655 & 4.557 & 43.90 & 21.16 & 3.010 & 0.1323\\
6.310$\times10^{-3}$ & 6.711 & 4.666 & 44.61 & 22.19 & 3.116 & 0.1390\\
7.079$\times10^{-3}$ & 6.767 & 4.772 & 45.34 & 23.23 & 3.229 & 0.1460\\
7.943$\times10^{-3}$ & 6.825 & 4.874 & 46.08 & 24.26 & 3.346 & 0.1534\\
8.913$\times10^{-3}$ & 6.883 & 4.973 & 46.83 & 25.27 & 3.469 & 0.1609\\
1.000$\times10^{-2}$ & 6.942 & 5.067 & 47.60 & 26.27 & 3.595 & 0.1686\\
1.122$\times10^{-2}$ & 7.001 & 5.158 & 48.38 & 27.24 & 3.725 & 0.1763\\
1.259$\times10^{-2}$ & 7.061 & 5.244 & 49.17 & 28.19 & 3.858 & 0.1839\\
1.413$\times10^{-2}$ & 7.120 & 5.326 & 49.96 & 29.11 & 3.991 & 0.1914\\
1.585$\times10^{-2}$ & 7.180 & 5.404 & 50.77 & 30.00 & 4.125 & 0.1986\\
1.778$\times10^{-2}$ & 7.240 & 5.480 & 51.58 & 30.87 & 4.258 & 0.2056\\
1.995$\times10^{-2}$ & 7.300 & 5.553 & 52.39 & 31.73 & 4.390 & 0.2125\\
2.239$\times10^{-2}$ & 7.360 & 5.627 & 53.22 & 32.61 & 4.521 & 0.2194\\
2.512$\times10^{-2}$ & 7.420 & 5.703 & 54.05 & 33.53 & 4.650 & 0.2266\\
2.818$\times10^{-2}$ & 7.481 & 5.787 & 54.90 & 34.55 & 4.780 & 0.2350\\
3.162$\times10^{-2}$ & 7.543 & 5.882 & 55.75 & 35.74 & 4.914 & 0.2455\\
3.548$\times10^{-2}$ & 7.607 & 5.996 & 56.63 & 37.19 & 5.055 & 0.2600\\
3.981$\times10^{-2}$ & 7.675 & 6.138 & 57.53 & 39.04 & 5.207 & 0.2816\\
4.467$\times10^{-2}$ & 7.747 & 6.317 & 58.45 & 41.47 & 5.375 & 0.3165\\
5.012$\times10^{-2}$ & 7.826 & 6.544 & 59.37 & 44.69 & 5.557 & 0.3784\\
5.623$\times10^{-2}$ & 7.917 & 6.824 & 60.27 & 48.97 & 5.734 & 0.5073\\
6.310$\times10^{-2}$ & 8.034 & 7.153 & 61.09 & 54.62 & 5.854 & 0.9048\\
7.079$\times10^{-2}$ & 8.226 & 7.492 & 61.77 & 62.02 & 5.769 & -22.99\\
7.943$\times10^{-2}$ & 8.602 & 7.752 & 62.32 & 71.77 & 5.097 & -0.5390\\
8.913$\times10^{-2}$ & 9.308 & 7.945 & 63.43 & 86.32 & 2.675 & -0.1168\\
 \hline
\end{tabular}
\end{center}
\end{table*}
\endgroup

\begingroup
\squeezetable
\begin{table*}
\caption{Ratio of the hybridization parameter $\lambda_C$ when the neutron superfluid density is that obtained in Ref.\cite{Chamel2012}, and the hydridization parameter $\lambda=v_{np}^2/(v_p^2-v_n^2)$ when the neutron superfluid density is equal to the neutron density outside nuclei.}
\label{tableChamel}
\begin{center}
\begin{tabular}{|c|c|}
  \hline
  $n [\texttt{fm}^{-3} ]$& $\lambda_C/\lambda$\\ \hline

3.162$\times10^{-4}$ & 0.6408\\
1.000$\times10^{-3}$ & 1.222\\
5.012$\times10^{-3}$ & 6.421\\
1.000$\times10^{-2}$ & 7.276\\
1.995$\times10^{-2}$ & 4.217\\
3.162$\times10^{-2}$ & 3.948\\
3.981$\times10^{-2}$ & 4.864\\
5.012$\times10^{-2}$ & 29.43\\
6.310$\times10^{-2}$ & -4.315\\
7.080$\times10^{-2}$ & 0.07204\\
7.943$\times10^{-2}$ & 2.619\\
 \hline
\end{tabular}
\end{center}
\end{table*}
\endgroup

\begin{figure}
\begin{center}
\includegraphics[width=9cm]{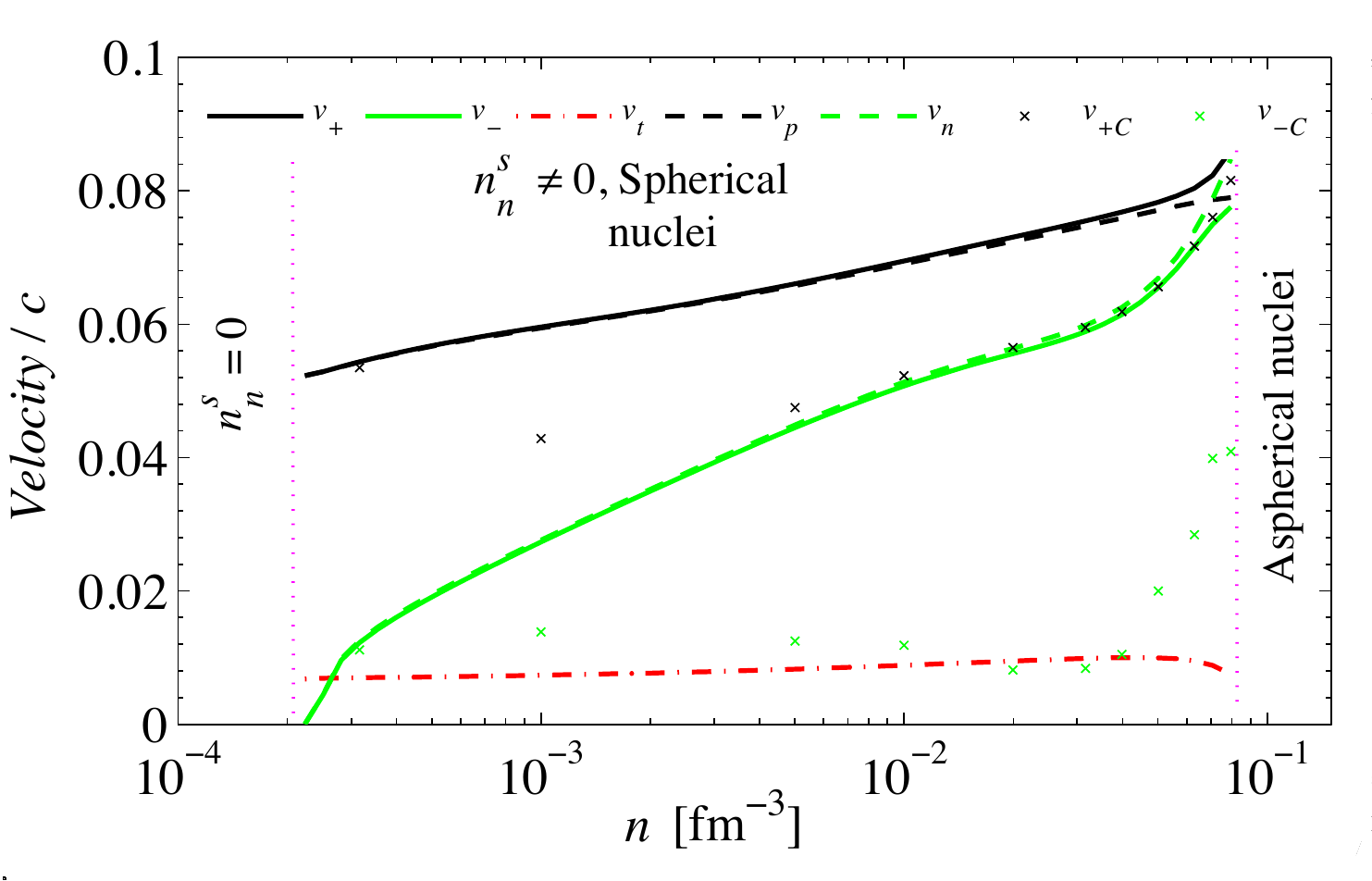}
\end{center}
\caption{Mode velocities in the inner crust.  The solid lines represent the results of the full calculations based on Eq.\ (\ref{vpm}) and Table I.  The lines correspond to the results for a neutron superfluid density equal to the density of neutrons between nuclei  with hybridization (full lines) and in the absence of hybridization ($v_{np}=0$, dashed lines).  The crosses show results for Chamel's values for the superfluid density \cite{Chamel2012}.}
\label{fig:modevelocities}
\end{figure}

In Table \ref{tab2} we show numerical results for $v_n^2$, $v_p^2$, $v_{np}^2$, $v_\pm^2$ calculated from Eq.\  (\ref{modevel}) and the thermodynamic derivatives given in Table \ref{tab1}.  For the neutron superfluid density, we took it to be the density of neutrons outside nuclei, $n_n^{\rm o}$. Plots of the mode velocities are given in Fig. \ref{fig:modevelocities}.    The quantity
\beq
\lambda=\frac{v_{np}^2}{v_p^2-v_n^2}.
\eeq
is a measure of the mixing of the uncoupled modes with velocities  $v_p$ and $v_n$.  In terms of this variable, the ratio of the amplitude of the uncoupled proton and neutron modes in the mode with velocity $v_-$  is
\beq
\tan \theta=  -\frac{2\lambda}{1+\sqrt{1+\lambda^2}}.
\eeq
The mixing is thus typically of order 10\% in the amplitude or 1\% in the probability except close to the crust--core boundary where the two uncoupled mode velocities cross.

In Ref.\ \cite{ChamelPageReddy} estimates of the hybridization of the two modes have been made with the neutron superfluid density being given by the results of Chamel's calculations \cite{Chamel2012}.  We have also made calculations with values of $n_n^s$ given in Ref. \cite{Chamel2012}, which for densities $n \approx (1-5)\times10^{-2} $ fm$^{-3}$ can be of order 0.1 $n_n^{\rm o}$ and the results are shown in Table \ref{tableChamel}.  The hybridization is generally much greater than for the choice $n_n^s=n_n^{\rm o}$, which will result in a considerably larger Landau damping of phonons in the neutron superfluid.   The four highest density values are close to the crust--core boundary, where the basic model of spherical nuclei fails, and thus the results cannot be regarded as physically meaningful.

\section{Microscopic theory}
We now consider the effective interaction between nuclei from a microscopic point of view.  Consider for the moment two impurities inserted into an initially uniform quantum liquid.  In the usual formulation of perturbation theory one works with the grand potential, $\Omega(\mu)=E-\mu \langle N\rangle$, where $E$ is the energy of the system, $\mu$ is the chemical potential and $N$ the number of particles.
Changes in $\Omega$ given in terms of linked graphs.  The single-impurity energy given  by all graphs in which a single impurity is coupled to the excitations of the liquid, and the two-impurity interaction is given by all linked graphs in which excitations of the liquid propagate between the two impurities.

\begin{figure}
\begin{center}
\includegraphics[width=8cm]{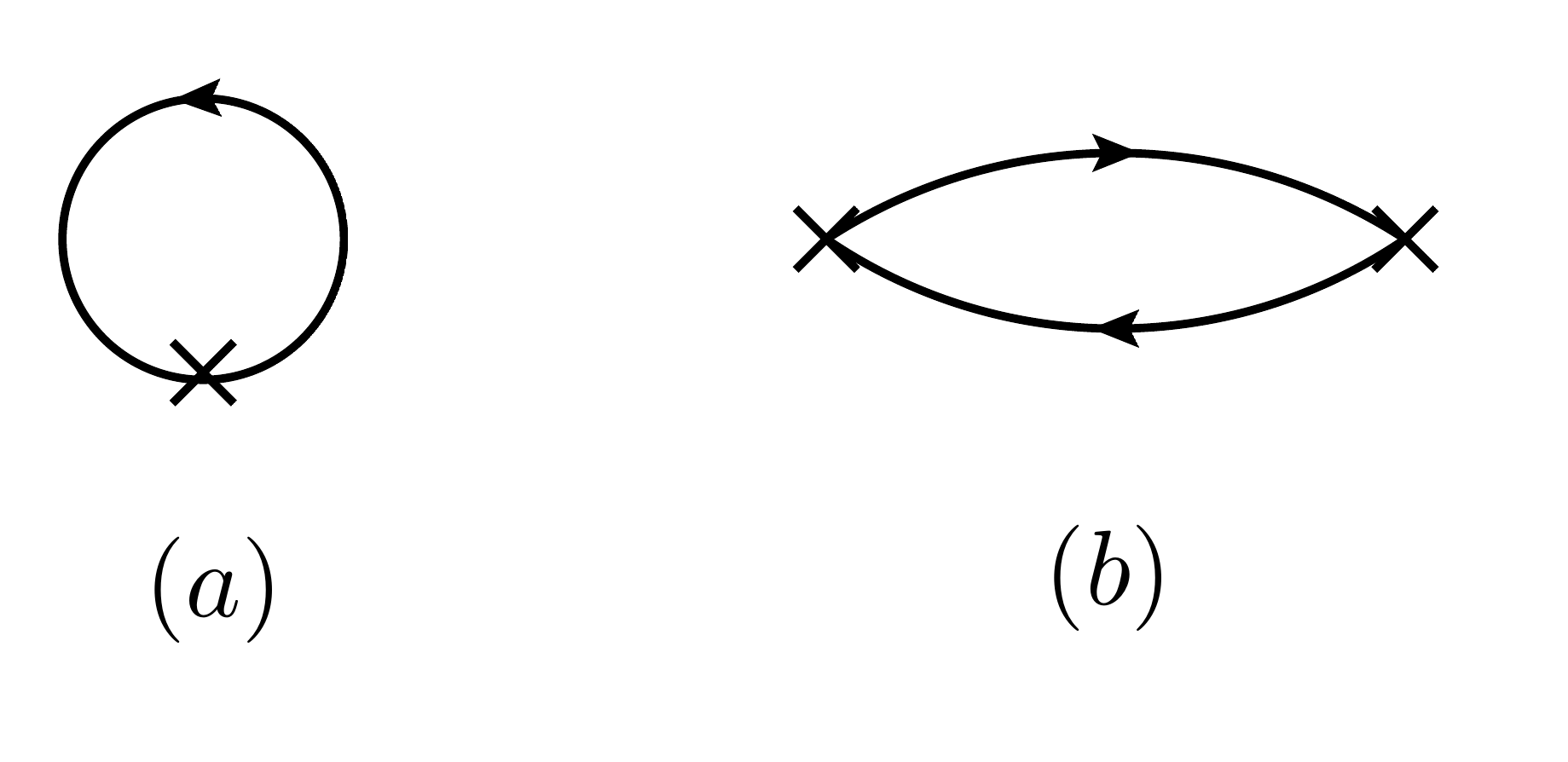}
\end{center}
\vspace{-2em}
\caption{Diagrams representing the simplest contributions to the energy of a single impurity, $(a)$, and to the interaction energy of two impurities in a normal Fermi gas, $(b)$. }
\label{Hartree+BubbleLabel}
\end{figure}

\begin{figure}
\begin{center}
\includegraphics[width=8cm]{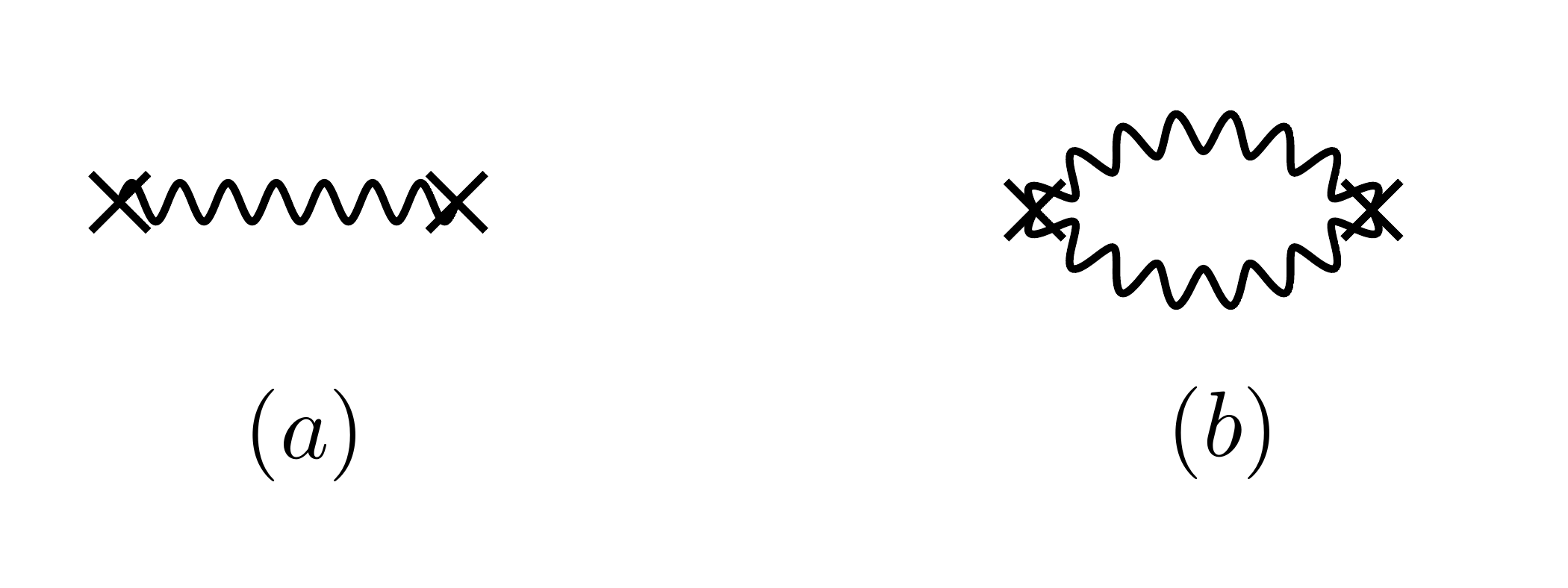}
\end{center}
\vspace{-2em}
\caption{Diagrams representing the contribution to the interaction energy between two static impurities in a quantum liquid due to one-phonon exchange, (a), and two-phonon exchange (b).  A cross denotes an impurity and the wavy line a phonon propagator in the quantum liquid.}
\label{1and2PhononExchange}
\end{figure}

\begin{figure}
\includegraphics[width=8cm]{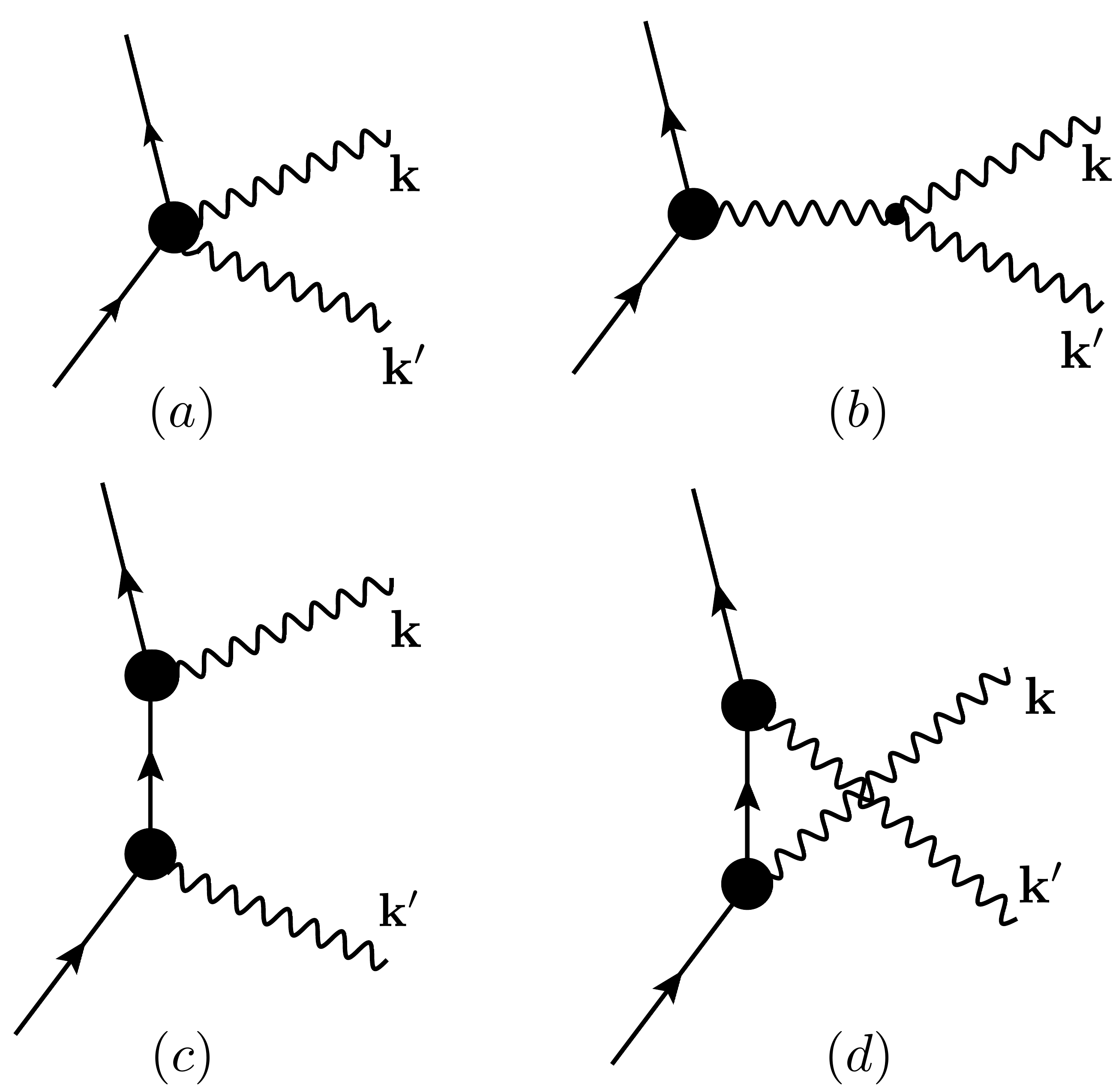}
\caption{Diagrams representing contributions to the vertex for coupling of a nucleus (solid lines) to two phonons (wavy lines).}
\label{TwoPhononEmission}
\end{figure}

For a normal Fermi system, the simplest contribution to the energy of a single impurity is the Hartree term, represented for a normal Fermi system by the graph shown in Fig.\ 2(a), and the simplest example for the interaction of two impurities is the graph shown in Fig.\ 2(b), which gives the analog for density perturbations of the Ruderman--Kittel interaction between spins.   More generally, interactions between the particles in the quantum liquid will give rise to more complicated contributions and the sum of connected graphs connecting the two impurities gives the extra energy required to add the second impurity minus the energy to add the first impurity, \emph{when the chemical potential of the quantum liquid is fixed}. Thus, for short-range interactions, this interaction energy tends to zero for large separations, unlike the direct and induced interactions in the thermodynamic formulation.

For definiteness, let us consider the interaction between two impurities in a superfluid Fermi liquid, for which the relevant long-wavelength degrees of freedom are (Bogoliubov--Anderson) phonons.  The simplest contribution to the interaction between impurities is represented by the graph shown in Fig. \ref{1and2PhononExchange}$(a)$, where the wavy line represents a density--density response function in the medium and a cross denotes an impurity.  In coordinate space this interaction is given by the Fourier transform of the density--density response function if one neglects the form factor for the scattering of a fermion by the impurity and, consequently, the spatial range of this term is of order the interparticle separation of the fermions.

There is a longer range part of the interaction due to the exchange of two or more phonons, as shown in Fig. 3$(b)$.  As explained in the context of helium mixtures \cite{BaymEbner}, the vertex for an impurity to scatter a phonon or create two phonons is made up of a number of contributions, as shown in Fig. 4, where we now indicate the impurity by a line rather than just a cross to indicate that the impurity may be mobile, as well as have internal degrees of freedom such as vibrational modes.   In the case of electromagnetism, when the exchanged excitations correspond to photons or the Coulomb interaction, such processes lead to the van der Waals and Casimir--Polder interactions \cite{CasimirPolder,Holstein}.

In general, the vertex is related to derivatives of the energy of a nucleus, its effective mass and the sound velocity with respect to the neutron density. These derivatives have not yet been evaluated for the inner crust, so we shall, for illustration, consider only the term in Fig. 4 (a) and neglect dependence of the effective mass on neutron density.  The energy of a nucleus to second order in changes in the neutron density may be written
\bea
\delta^2E= \frac12 g(\delta n_n)^2,
\eea
where
\bea
g=\frac{\partial^2
\epsilon_N}{\partial n_n^2},
\eea
where $\epsilon_N$ is energy to add a single nucleus to the neutron fluid.
The neutron density operator is related to phonon creation and annihilation operators by the equation
 \bea
 \delta n_n({\bf r})=\sum_{\bf k}\left(\frac{n_n k}{2  m s \,{\cal V}}     \right)^{1/2}(b_{\bf k}\rme^{\rmi {\bf k}\cdot {\bf r}}  +b_{\bf k}^\dagger\rme^{-\rmi {\bf k}\cdot {\bf r}}),
 \label{densityfield}
\eea
where $s$ is the sound velocity in the neutron superfluid and $n_n$ the neutron number density. For simplicity we work in units for which $\hbar=1$. The interaction energy due to two-phonon exchange between two nuclei separated by distance $r$  by second-order perturbation theory and has the form
 \begin{align}
V_2(r)=- g^2\left(\frac{n_n}{2m s}\right)^2\int \frac{d^3k}{(2\pi)^3}  \frac{d^3k'}{(2\pi)^3} \frac{kk'}{s(k+k')}\rme^{-\rmi({\bf k}+{\bf k'})\cdot {\bf r}},\nonumber \\
 \end{align}
where the factor $kk'$ comes from Eq.\ (\ref{densityfield}) and the $s(k+k')$ is the energy denominator.  On performing the angular integrations, this equation becomes
  \begin{align}
V_2(r)=\hspace{20em} \nonumber \\- \frac{g^2}{r^2}\left(\frac{n_n}{4 \pi^2 m s}\right)^2\int_0^\infty   \int_0^\infty dk  dk' \frac{k^2k'^2}{k+k'}\sin{kr}\sin{k'r}.\nonumber \\
 \end{align}
With the use of the identity $\int_0^\infty d\sigma \rme^{-\sigma (k+k')}=1/(k+k')$ and rescaling $k$ and $\sigma$ one finds
 \begin{align}
V_2(r)=
- \frac{g^2}{r^7}\left(\frac{n_n}{4\pi^2m s}\right)^2\int_0^\infty d\sigma\left[\int_0^\infty {dk}\rme^{-\sigma k} k^2\sin k\right]^2.  \nonumber \\
\hspace{-5em}
\end{align}
The integral in square brackets is $(-2 + 6 \sigma^2)/(1 + \sigma^2)^3$ and
\be
\int_0^\infty d\sigma\left[ \frac{-2 + 6 \sigma^2}{(1 + \sigma^2)^3}    \right]^2 =\frac{3\pi}{8},
\ee
from which it follows that
\be
V_2(r)=- \frac{3}{128\pi^3}\frac{ g^2n_n^2\hbar}{m^2 s^3}\frac{1}{r^7},
\ee
where we have reinserted $\hbar$.   This of the Casimir--Polder form.

To obtain a rough estimate of $g$ we adopt a model of the nucleus as a hard sphere of volume $V_N$.  The energy to add one nucleus to a uniform neutron fluid is then $V_N P(n_n^{\rm o})$ and therefore
\be
g= V_N \frac{\partial^2 P(n_n^{\rm o})}{\partial (n_n^{\rm o})^2},
\ee
where $P(n_n^{\rm o})$ is the pressure of the neutron fluid.  Since $dP=n_n d
\mu_n$ at zero temperature, it follows that
\be
\frac{\partial^2 P(n_n^{\rm o})}{\partial (n_n^{\rm o})^2}=\frac{\partial }{\partial n_n^{\rm o}}\left( n_n^{\rm o}   \frac{\partial \mu(n_n^{\rm o})}{\partial n_n^{\rm o}}        \right)
\ee
If the neutron fluid is treated as a noninteracting Fermi gas,  $P$ varies as $(n_n^{\rm o})^{5/3}$ and $\mu_n$ as $(n_n^{\rm o})^{2/3}$, and it follows that
\be
V_2(r)=- \frac{1}{96\pi^3}\frac{V_N^2\hbar s}{r^7}.
\ee

For nuclei in the inner crust, the nuclear radius $r_N$ is $\sim 6$ fm and therefore when the two nuclei touch ($r=2r_N$), $V_2$ is on the keV scale.  This is small compared with the Coulomb energy of two nuclei at this separation, $Z^2e^2/(2r_N)\sim 10^2$ MeV.  This indicates that it is a good first approximation to neglect the Van der Waals interaction compared with the Coulomb interaction.

In this calculation we have considered the interaction between two nuclei in an otherwise uniform neutron fluid, which should be a good first approximation at densities not too far above that for neutron drip.  However, closer to the crust--core boundary, it is important to take into account interactions between three or more nuclei, which will give rise to band-structure effects on the exchanged phonons.

\section{Concluding remarks}

One of the main conclusions of this article is that, when the Coulomb interaction between nuclei is neglected and in the static limit,  the induced interaction between two nuclei in the inner crust of a neutron star is exactly cancelled by the direct interaction in any model in which the energy of a nucleus is a function only of the density of the surrounding neutrons.
The cancellation is rather general, and in Appendix A we have demonstrated it explicitly for a model in which surface effects are neglected.

This cancellation points to the need to investigate in greater detail the spatial dependence of the interaction between nuclei, and as a first step we have shown that at large distances the total interaction between two nuclei is attractive and of the Casimir--Polder form, $\sim 1/r^7$, due to exchange of two phonons.  At smaller separations there are contributions to the interaction due to exchange of higher numbers of phonons.   More generally, when the energy exchange between the two nuclei is nonzero, the cancellation of the two contributions will be incomplete because of the nonzero time required for the neutron density to respond to a change in the configuration of the nuclei.  Our rough estimates of the size of the Casimir--Polder interaction suggest that it is considerably smaller than the direct Coulomb interaction between nuclei and therefore it seems unlikely that it can destabilize the lattice, as suggested in Ref.\ \cite{KP2}.

Our calculations of collective modes show that results for the degree of hybridization of lattice phonons and phonons in the neutron superfluid is very sensitive to the neutron superfluid density.  This implies that estimates of Landau damping of the phonons in the neutron superfluid also depend on the neutron superfluid density.  A challenge for the future is to calculate the neutron superfluid density taking into account both band-structure effects and pairing.

\section{Acknowledgments}

We are grateful to Michael Schecter for illuminating discussions of the Casimir--Polder interaction.  Part of this work was done while the authors enjoyed the hospitality of  ECT* and while CJP was at the Aspen Center for Physics, which is supported by National Science Foundation grant PHY-1066293.    This work was also supported in
part by the Russian Fund for Basic Research grant 31 16-32-60023/15 and by NewCompStar, COST Action MP1304.

\appendix
\section{Thermodynamic derivatives for the bulk equilibrium approximation}
Here we give an explicit example of the calculation of the derivatives $E_{ij}$ for the case of two coexisting bulk phases, one consisting of pure neutron matter with density $n_n^\rmo$ and pressure $P^\rmo$ and the other with nuclear matter with neutron density $n_n^\rmi$, proton density $n_p^\rmi$ and pressure $P^\rmi$.
In this calculation we assume electrons to be absent and neglect the Coulomb interaction.
We denote the fraction of space occupied by the proton-rich phase by $u$. For the two phases to be in equilibrium, the neutron chemical potential $\mu_n$ in the two phases must be equal and the pressures must be equal.  When the densities of neutrons and protons in the two phases change, the condition for the pressures of the two phases to remain equal is
\be
dP^\rmi=dP^\rmo
\ee
or
\be
n_n^\rmi \delta \mu_n +n_p^\rmi \delta \mu_p^\rmi=n_n^\rmo \delta \mu_n,
\ee
where we have already used the condition that in equilibrium the neutron chemical potentials inside and outside nuclei must be equal.
It therefore follows that along the coexistence line,
\be
\delta \mu_n=-\frac{n_p^\rmi}{n_n^\rmi-n_n^\rmo}\delta \mu_p^\rmi.
\label{delta_mu_s}
\ee
The change in the total neutron density is given by
\be
\delta n_n=u\delta n_n^\rmi+(1-u)\delta n_n^\rmo +(n_n^\rmi-n_n^\rmo)\delta u,
\label{deltannbulkequi}
\ee
and that of the proton density by
\be
\delta n_p=u\delta n_p^\rmi+n_p^\rmi \delta u.
\ee
At fixed $n_p$ it therefore follows that
\be
\delta u=-\frac{u}{n_p^\rmi }\delta n_p^\rmi,
\ee
and, from Eq.\ (\ref{deltannbulkequi}),
\begin{align}
\delta n_n &=u(\delta n_n^\rmi-\nu\delta n_p^\rmi)+(1-u)\delta n_n^\rmo\,.
\label{deltanndeltamun}
\end{align}
 To linear order, one can write
 \begin{align}
 \delta n_j^\rmi = -\Omega^\rmi_{jk} \delta \mu_k\,,
 \label{dn_dmu}
 \end{align}
where $\Omega_{ij}^\rmi=\partial^2 \Omega^\rmi/\partial \mu_i\partial \mu_j$, $\Omega$ being the grand potential of uniform matter per unit volume.
 From Eq.~(\ref{deltanndeltamun}) it therefore follows that
\begin{align}
E_{nn}&=\left.\frac{\partial \mu_n}{\partial n_n}\right|_{n_p}\nonumber\\
&=-\left[ u(\Omega_{nn}^\rmi -2\nu\Omega_{np}^\rmi+ \nu^2\Omega_{pp}^\rmi) +(1-u)\Omega_{nn}^\rmo          \right]^{-1}.
\label{EnnApp}
\end{align}
The matrix $\Omega_{ij}^\rmi$ is the negative of the inverse of $E^\rmi_{ij}$, where the subscript $\rmi$ indicates that the quantity is to be evaluated for matter inside nuclei.

Since the ratio of changes in chemical potentials is given by
Eq.~(\ref{delta_mu_s}), it follows that
\begin{align}
E_{np}=\left.\frac{\partial \mu_p}{\partial n_n}\right|_{n_p}=-\nu E_{nn}.
\label{mixedderivApp}
\end{align}
Next we calculate the changes in the proton and neutron chemical potentials when the number density of neutrons is held fixed.  The condition for constancy of the neutron density is
\be
\delta n_n=u\delta n_n^\rmi+(1-u)\delta n_n^\rmo +(n_n^\rmi-n_n^\rmo)\delta u=0,
\ee
and therefore the change in the proton density is given by
\begin{align}
\delta n_p&=u\delta n_p^\rmi +n_p^\rmi \delta u \nonumber \\
&=u\left( \delta n_p^\rmi -\frac{\delta n_n^\rmi}{\nu}  \right)-(1-u)\frac{\delta n_n^\rmo}{\nu}.
\end{align}
Since the changes in the number densities are given by Eq.~(\ref{dn_dmu}) and the changes in the chemical potentials are related by Eq.~(\ref{delta_mu_s}), one finds
\be
E_{pp}^{nuc}=\nu^2 E_{nn}\,.
\label{EppApp}
\ee
In addition, the result for the mixed derivative obtained from calculating ${\partial \mu_n}/{\partial n_p}$ agrees with Eq.\ (\ref{mixedderivApp}), as it must on general grounds. From Eqs.~(\ref{EnnApp}), (\ref{mixedderivApp}) and (\ref{EppApp}) one sees that $E_{nn}E_{pp}^{nuc}-E_{np}^2=0$.


\begin{thebibliography}{99}
%
\bibitem{Epstein}  R. I. Epstein, {Astrophys. J. }{\bf 333},
880 (1988).

\bibitem{ChamelPageReddy}
N. Chamel, D. Page and S. Reddy, { Phys. Rev. C} {\bf 87}, 035803 (2013)).
\bibitem{KP1} D. Kobyakov and C. J. Pethick, { Phys. Rev. C} {\bf 87}, 055803 (2013).
\bibitem{KP2} D. Kobyakov and C. J. Pethick, {Phys. Rev. Lett.} {\bf 112}, 112504 (2014).

\bibitem{BBP1}  J.\ Bardeen, G.\ Baym, and D.\ Pines, Phys.\ Rev.\ {\bf 156}, 207 (1967).

\bibitem{BabuBrown} See, e.g., S. Babu and G. E. Brown, {Ann. Phys.} {\bf 78}, 1 (1973).

\bibitem{LattimerSwesty}  J.~M.~Lattimer and F.~D.~Swesty, Nucl.~Phys.~A {\bf 535}, 331 (1991) and the website\\ www.astro.sunysb.edu/dswesty/lseos.html.

\bibitem{LandLHydro} L.\ D.\ Landau and E.\ M.\ Lifshitz, \emph{Fluid Mechanics}, 2nd ed., (Pergamon, Oxford, 1987), \S 139.

\bibitem{CJPChamelReddy}  C. J. Pethick, N. Chamel, and S. Reddy, Prog.\ Theor.\ Phys.\ Suppl.\ {\bf 186}, 9 (2010).

\bibitem{Schecter}  In the context of low-dimensional systems the importance of the Casimir--Polder interaction has been stressed by M. Schecter and A. Kamenev, Phys.\ Rev.\ Lett.\
\textbf{112}, 155301 (2014).

\bibitem{RavenhallBP} D. G. Ravenhall, C. D. Bennett, and C. J. Pethick,  Phys.\ Rev.\ Lett.\
\textbf{28}, 978 (1972).
\bibitem{NegeleVautherin}  J. W. Negele and D. Vautherin, Nucl. Phys.  A \textbf{207}, 298 (1973).

\bibitem{DouchinHaensel}F. Douchin and P. Haensel, Astron. and Astrophys. {\bf 380}, 151 (2001).

\bibitem{GrillMargueronSandulescu} F. Grill, J. Margueron, and N. Sandulescu, Phys.\ Rev.\ C \textbf{84}, 065801 (2011).
\bibitem{PearsonChamel}J. M. Pearson, N. Chamel, S. Goriely, and C. Ducoin, Phys.\ Rev.\ C \textbf{85}, 065803 (2012).
\bibitem{BaldoBurgio} B. K. Sharma, M. Centelles, X. Vinas, M. Baldo, G. F. Burgio, Astron. and Astrophys. {\bf584}, A103 (2015).

\bibitem{1+alpha}In the theory of helium mixtures, the quantity $n_1v_2$ is denoted by $1+\alpha$, where $\alpha$ is the relative excess volume occupied by a $^3$He atom compared with a $^4$He atom \cite{BBP1}.

\bibitem{mue included in Epp}  From the work of Ref.\ \cite{LattimerSwesty} one can calculate $E_{pp}$ for electrically neutral matter.  The quantity $\partial E/\partial n_p$ for such matter contains a term $\mu_e$, since when a proton is added to matter, an electron must also be added to maintain charge neutrality.  However, although $\mu_e$ can be as large as $\sim 60$ MeV, it does not depend on $n_n$ when $n_p$ is fixed and therefore does not contribute to $E_{np}$.




\bibitem{Carter} See, e.g., B. Carter and I. M. Khalatnikov, { Phys. Rev. D} {\bf 45}, 4536 (1992).

\bibitem{Kobyakov_etal} D. Kobyakov, C. J. Pethick, S. Reddy, and A. Schwenk, (in preparation).

\bibitem{Releffects}For simplicity, we replace mass densities by rest-mass densities.  This approximation is most serious for the electrons, since their effective mass is $\mu_e/c^2$, which can amount to a $\sim$10\% addition to the normal density of the protons at the inner edge of the crust, where the electron chemical potential approaches 100 MeV.


\bibitem{Chamel2012} N. Chamel, Phys. Rev. C {\bf 85}, 035801 (2012).


\bibitem{BaymEbner}  G. Baym and C. E. Ebner, Phys. Rev. {\bf 164}, 235 (1967).

\bibitem{CasimirPolder} H. B. G. Casimir and D. Polder,  Phys. Rev.  {\bf 73}, 360 (1948).

\bibitem{Holstein}  For a pedagogical review of the van der Waals and Casimir--Polder interactions see B.  Holstein, Am. J. Phys. {\bf 69}, 441 (2001).



\end{thebibliography}
\end{document}